\documentclass[fleqn,twoside]{article}
\usepackage{espcrc2}
\usepackage{graphicx}
\def\Tr   {\mathop{\hbox{Tr}}}
\def\det   {\mathop{\hbox{det}}}

\newcommand{\AmS}{{\protect\the\textfont2
  A\kern-.1667em\lower.5ex\hbox{M}\kern-.125emS}}

\title{{ Improving the Partial-Global Stochastic Metropolis Update
for Dynamical Smeared Link Fermions} }

\author{A. Hasenfratz\address[MCSD]{Department of Physics, University of Colorado, 
        Campus Box 390, Boulder, CO 80309, USA}
        and
        A. Alexandru\addressmark[MCSD]
       }

\begin{document}

\begin{abstract}
We discuss several methods that improve the partial-global stochastic Metropolis
(PGSM) algorithm for smeared link staggered fermions. We present autocorrelation
time measurements and argue that this update is feasible even on reasonably
large lattices. 
\vspace{1pc}
\end{abstract}
\vskip -0.4cm

% typeset front matter (including abstract)
\maketitle

Smeared link actions have gained popularity in recent years. These
actions have many desirable features: improved flavor symmetry with
staggered fermions, smaller perturbative matching factors, better
topology, etc \cite{proc_ah}. What hinders the use of these smeared actions is the
difficulty of dynamical simulations with standard molecular dynamics type algorithms.
 Recently it was suggested that even projected
smeared link actions can be effectively simulated with partial-global
heatbath or over-relaxation updates that are combined with a Metropolis type
accept-reject step \cite{PGSM}. In this paper we summarize some of the improvements
that make this partial-global stochastic Metropolis (PGSM) update truly
effective.

\section{THE SMEARED LINK ACTION AND THE PARTIAL-GLOBAL UPDATE }

We consider a smeared link action of the form\begin{equation}
\label{full_action}
S=S_{g}(U)+\bar{S}_{g}(V)+S_{f}(V),
\end{equation}
 where \( S_{g}(U) \) and \( \bar{S}_{g}(V) \) are gauge actions
depending on the thin links \( \{U\} \) and smeared links \( \{V\} \),
respectively, and \( S_{f} \) is the fermionic action depending on
the smeared links only. The smeared links are constructed deterministically.
Here we use HYP blocking \cite{coloquench}.
\( S_{g}(U) \) is an arbitrary gauge action,
and \( \bar{S}_{g}(V) \) contains 4 and 6-link loops of the smeared links.
We choose the  coefficients of \( \bar{S}_{g}(V) \) 
to improve computational efficiency.
The fermionic action describing \( n_{f} \) degenerate flavors with
staggered fermions is \( S_{f}(V)=-\frac{n_{f}}{4}\ln \, \det \, \Omega (V) \)
with \( \Omega (V)=(M^{\dagger }(V)M(V)) \) defined
on even sites only. 

We use the partial-global stochastic updating (PGSM) algorithm to
simulate this system \cite{PGSM}. 
The PGSM algorithm is a variation of the pseudo-fermion
method and satisfies the detailed balance equation. First a subset
of the thin links \( \{U\} \) are updated and a new thin gauge link
configuration \( \{U'\} \) is proposed such that the transition probability
\( p(U,U') \) of the update satisfies detailed balance with \( S_{g}(U) \)
. The proposed configuration is accepted with the probability \begin{equation}
\label{Pstoch}
P_{\rm {acc}}=\rm {min}\{1,e^{-\Delta \bar{S}_{g}}e^{-\xi ^{*}[A ^{n_f/4}-1]\xi }\},
\end{equation}
 where \(\Delta \bar{S}_{g}=\bar{S}_{g}(V')-\bar{S}_{g}(V) \), 
the matrix $A=\Omega^{'-1/2}\Omega\Omega^{'-1/2}$,  and 
the stochastic vector \( \xi  \) is generated with Gaussian
distribution, \( \rm {exp}(-\xi ^{*}\xi ) \).

\section{IMPROVING THE PGSM UPDATE}

The PGSM algorithm averages the stochastic estimator of the fermionic
determinant together with the gauge configuration ensemble average
and it can be efficient only if the standard deviation of the stochastic
estimator is small. 
The standard deviation of \( \det  A ^{-1}\)  can be written
as {\begin{eqnarray}
\sigma _{A}^{2} & = & <e^{-2\xi ^{*}[A-1]\xi }>-<e^{-\xi ^{*}[A-1]\xi }>^{2}\nonumber \\
 & = & \det (2A-1)^{-1}-\det (A)^{-2}.\nonumber \label{standard_dev} 
\end{eqnarray}
} \( \sigma _{A} \) diverges if even one of the eigenvalues of \( A \)
is less than 1/2, which is likely to occur if \( \Omega  \) and \( \Omega ' \)
differ significantly \cite{det}. We consider two separate methods
to solve this problem.

\textbf{1. Reduction:} The fermionic matrix reduction \cite{PGSM,det}
removes the most UV part of the fermionic operator by defining a reduced
matrix \( \Omega _{r}=\Omega e^{-2f(\Omega )} \)
 and rewriting the fermionic action as \begin{equation}
\label{S_f_mod}
S_{f}(V)=-\frac{n_{f}}{4}\Big (\ln \, \det \, \Omega _{r}(V)+2\Tr f(\Omega )\Big ).
\end{equation}
 If the function \( f \) is a polynomial of \( \Omega  \), 
 \( \Tr f(\Omega ) \) can be evaluated exactly and only the determinant
of the reduced matrix \( \Omega _{r} \) has to be calculated stochastically.
The parameters of \( f \) are chosen such as to minimize the eigenvalue
spread of  \( \Omega _{r} \). With
\( f \) a cubic polynomial the conditioning number of the fermionic
matrix can be reduced by about a factor of 30.

\textbf{2. Determinant breakup}: Writing the fermionic determinant in the
form 
\( \det  A ^{-1}= (\det A ^{1/n_b} )^{-n_b}\) 
 suggests that the stochastic part of the estimator can be evaluated
using \( n_{b}n_{f}/4 \) independent Gaussian vectors with the replacement \begin{equation}
\label{P_acc_mod}
e^{-\xi^{*} (A^{n_f/4}-1)\xi } \to  e^{-\sum ^{n_{b}n_{f}/4}_{i=1}\xi _{i}^{*}[A^{1/n_{b}}-1]\xi _{i}}.
\end{equation}
The standard deviation of the stochastic estimator is now finite as
long as the matrix \( \Omega _{r}^{-1/2}(V')\Omega _{r}(V)\Omega _{r}^{-1/2}(V') \) has no
eigenvalue smaller than \( 1/2^{n_{b}} \), a condition 
that can always be satisfied \cite{det}. 

\begin{figure}
%\vskip -1.2cm
\includegraphics[width=7cm,height=8cm]{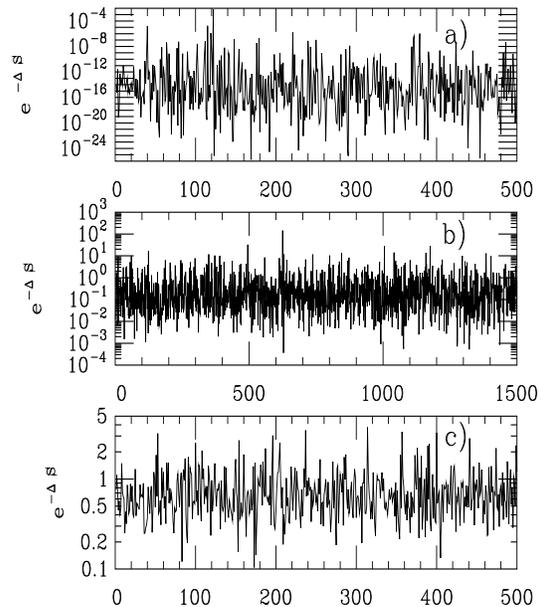}
\vskip -0.9cm
\caption{The stochastic estimator  on a given fermionic matrix \( A \) with a) no improvement; b) fermionic reduction but no determinant breakup; c) fermionic reduction and \( n_b=8 \) determinant breakup.}
\vskip -0.6cm
\label{stoch_est}
\end{figure}

The effectiveness of the fermionic reduction and determinant breakup
is illustrated in figure~\ref{stoch_est} where we show the stochastic
estimator on a specific \( A\)
matrix using no improvement, with fermionic reduction but no determinant
breakup, and with fermionic reduction and \( n_{b}=8 \) determinant breakup. Observe the 16
orders of magnitude difference between between figure \ref{stoch_est}a
and \ref{stoch_est}c.
\section{TUNING THE ACTION PARAMETERS}

Improving the stochastic estimator ensures that the acceptance rate
of the PGSM update is close to the theoretical maximum, the
exact ratio of the determinants. But the acceptance rate will be large
only if the exact ratio is large. That will be the case if the configurations
proposed by the pure gauge action \( S_{g}(U) \) are close to the
dynamical configurations. Chosing \( S_{g}(U) \) such that it matches
the lattice spacing and/or short distance fluctuations of the expected
dynamical configurations can maximize the acceptance rate. Such a
matching requires the shift of the pure gauge coupling towards the
appropriate quenched value. This shift can be compensated by the \( \bar{S}_{g}(V) \)
term. \( \bar{S}_{g}(V) \) is taken into account in the accept-reject
step but since it depends on the smeared links only, it fluctuate
much less than \( S_{g}(U) \) and does not affect the acceptance
rate significantly \cite{auto}.

\section{THE EFFECTIVENESS OF THE PGSM UPDATE}

The effectiveness of the PGSM algorithm depends on both the partial-global
heat bath update and the stochastic estimator. The autocorrelation
time of the simulation is the larger of the autocorrelation times
of the heat bath and stochastic steps. We have measured the integrated
autocorrelation time of the plaquette as the function of the number
of links updated at one heat bath step, \( t_{HB} \) \cite{auto}.
The autocorrelation time of the pure gauge partial-global
heat bath update can be expressed as  
\begin{equation}
\label{tau_pg}
\tau _{pg}=\tau _{HB}\frac{4V}{t_{HB}r_{a}},
\end{equation}
 where \( \tau _{HB} \) is the autocorrelation time of the 
pure gauge heat bath update, \( V \) is the volume of the lattice,
and \( r_a \) is the acceptance rate.
\begin{figure}
\includegraphics[width = 6.5cm, height = 6.5cm, angle = -90]{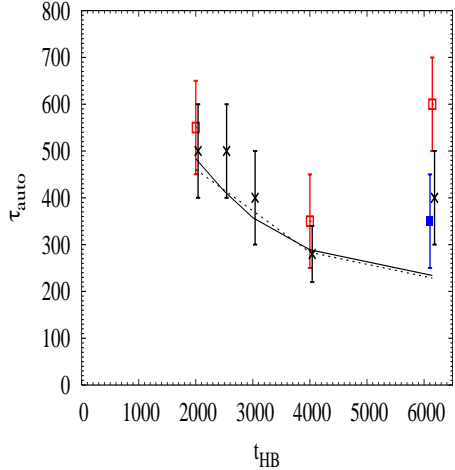}
\vskip -0.9cm
\caption{Autocorrelation of the \protect\( n_{f}=2\protect \) runs. Open
squares: \protect\( n_{b}=8\protect \), \protect\( am=0.10\protect \),
 filled squares: \protect\( n_{b}=12\protect \), \protect\( am=0.10,\protect \)
crosses: \protect\( n_{b}=12\protect \), \protect\( am=0.04\protect \)
runs. 
The lines corresponds to the expected autocorrelation based on the
heatbath update alone. 
}
\vskip -0.7cm
\label{tau_nf2}
\end{figure}
All our simulations were carried out on \( 8^{3}\times 24 \) lattices
at lattice spacing \( a\sim 0.17 \)~fm with \( n_f=2 \) fermion flavors. We considered two quark mass
values, \( am=0.10 \) where \( m_{\pi }/m_{\rho }\sim 0.71 \), and
\( am=0.04 \) where \( m_{\pi }/m_{\rho }\sim 0.54 \). In figure \ref{tau_nf2}
we plot \( \tau _{\rm {auto}} \) for the different runs showing the 
dependence on the quark mass, determinant breakup, and \( t_{HB} \).

\begin{figure}
\includegraphics[width=6.5cm, height=6.5cm, angle=-90]{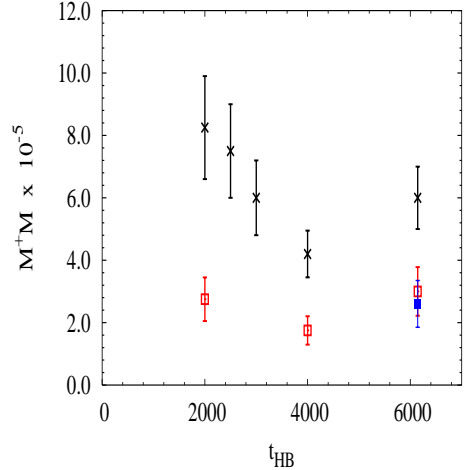}
\vskip -0.8cm

\caption{Cost of creating independent configurations, measured in fermionic
matrix multiplies. Notation is the same as in figure \ref{tau_nf2}.}
\vskip -0.7cm
\label{M+M_cost}
\end{figure}

We conclude this paper by translating the autocorrelation time measurements
to computer time requirements. In figure \ref{M+M_cost} we plot the
computer time measured in fermionic matrix \( M^{\dagger }M \) multiplies
to create configurations 
that are separated by \( 2\tau _{\rm {auto}} \) update steps. Using
thin link fermions with standard small step size algorithms, it takes
about \( 8\times 10^{5} \) \( M^{\dagger }M \) multiplies to
create independent configurations with \( am=0.10 \). On these 10~fm\( ^{4} \)
lattices it is actually faster to create HYP smeared configurations
with PGSM than thin link ones with a small step size algorithm. 
The PGSM algorithm
scales with the square of the lattice volume. To repeat the above
measurements on 100~fm\( ^{4} \) lattices would increase the computer
time by 100. A factor of 10 increase is due to the increased volume,
and the other factor of 10 is due to  the increased autocorrelation
time. Even on 100~fm\(^4\) volume the PGSM simulation with \( am =0.10 \) 
will be only a few times slower than thin link fermion simulations. 
Simulation cost at smaller lattice spacing but same physical
volume increases only linearly with the lattice volume as the number
of links that can be effectively touched increases accordingly.

\end{document}